\documentclass[doublecol]{epl2}

\usepackage{graphicx}% Include figure files
\begin{document}

%Title of paper
\title{Capacitor physics in ultra-near-field heat transfer}
\author{Jian-Sheng Wang \and Jiebin Peng}
\shortauthor{J.-S. Wang \& J. Peng}
%\email[]{phywjs@nus.edu.sg}
%\homepage[]{http://staff.science.nus.edu.sg/~phywjs}
%\thanks{}
%\altaffiliation{}
\institute{Department of Physics, National University of Singapore, Singapore 117551, Republic of Singapore}

\date{1 February 2017}


\abstract{
Using the nonequilibrium Green's function (NEGF) formalism, we propose a microscopic theory for near-field 
heat transfer between charged metal plates focusing on the Coulomb interactions.  
Tight-binding models for the electrons are coupled to the electromagnetic field continuum through a 
scalar potential. Our approach differs from the established ones based on Rytov fluctuational electrodynamics, which 
deals with the transverse radiative field and vector potential. For a two quantum-dot model a new length scale 
emerges below which the heat current exhibits great enhancement.  This length scale is related to the physics 
of parallel plate capacitors.  At long distances $d$, the energy flux decreases as $1/d^2$.
}

%\maketitle must follow title, authors, abstract, \pacs, and \keywords
\maketitle

% introduction
The thermal radiation in a cavity can be well-described by Planck's theory of black-body radiation \cite{Planck_book} - a great achievement of twentieth century physics, which started
the quantum physics revolution.  Two plates at temperatures $T_0$ and $T_1$ will transfer radiative heat at a rate proportional to $T_0^4 - T_1^4$ in the black-body limit, following the
Stefan-Boltzmann law. In the 70s both theoretical \cite{PvH} and 
experimental \cite{hargreaves69,domoto70,Ottens11} work have indicated corrections to the far field prediction when the distances between the plates are comparable to the thermal wavelengths of the electromagnetic fields. Near-field effects can be as large as a thousand fold that of the far field results \cite{volokitin07,shen09,Song-rev15}.  

Most recently, due to great progress in technology and precision measurements, much closer proximity is possible, on the scale of nanometers, or near contact. 
The near-field enhancement reported experimentally 
in Refs.~\cite{kimsong2015,song2016}
is consistent with the established theory.
However, the result in 
\cite{Kloppstech2015} is much too large to be explained by  
existing theories.  Are there other mechanisms for the near-field effect? 
%in comparison to black-body values, 
%while other experiments report results  .

Polder and van Hove (PvH) \cite{PvH} were the first to give a quantitative theory of near-field radiation using the Rytov formulation of fluctuating electromagnetic fields \cite{Rytov,bimonte16}.  The current-current correlation is assumed to follow the equilibrium fluctuation-dissipation theorem.  The average value of the Poynting vector is computed from the solution of macroscopic Maxwell equations.   In this picture, the near-field contribution is largely due to evanescent modes which are absent in the far field.  
A quantum electrodynamics treatment with linear media and NEGF reproduces the PvH theory \cite{Janowicz03}.

The large near-field effect is recently explained by phonon tunneling or surface
phonon polaritons 
\cite{mahanAPL11,xiong14,chiloyan14}.   In these work, although the Coulomb interaction is  dealt with indirectly, 
it is still treated as dipole-dipole fluctuations in a charge-neutral system.  We advocate that charge-charge fluctuations are important features
at short distances. 
The aim of this letter is to propose such a fundamental theory. 
We begin with
a tight-binding model of the electrons, as in a metal, for example, and couple it
to the scalar field in a quantized form.  While the electrons are on discrete lattices, the electromagnetic field 
is continuous and permeates the whole space.   Due to the scalar field nature, we need to use 
Lorentz gauge quantization \cite{tannoudji-book}, and then take the limit of the speed of light $c$ going to 
infinity in order to be consistent with the gauge condition. 

To demonstrate the basic idea, we define a toy
model for a nano-sized capacitor consisting of two quantum dots and a one-dimensional (1D) scalar field, mediating 
the Coulomb interactions of the charges.  
The same formulation can be applied to more realistic models, such as two graphene sheets. 
Figure 1 illustrates the model.

\begin{figure}
\includegraphics[width=0.95\columnwidth]{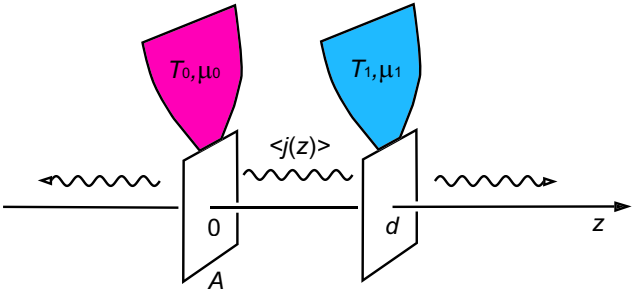}%
\caption{Schematic diagram of the 1D quantum-dot model.}
\end{figure}
%% section two dots model
We imagine a nanoscale parallel plate capacitor for which each of the plates can have a charge of 0 or 
$-Q$.
The plates located at $z=0$ and $d$ are connected to their respective electron baths so that their charges can fluctuate.  
The baths are necessary ingredient so that infinite amount of energy can be transferred from one dot to the other given
time.  The couplings of the quantum dots with the baths are bilinear in the system and bath fermion operators; 
their effects are captured by the self-energies of the baths \cite{wingreen-meir94}. 
We ignore the vector potential and consider
the scalar potential $\phi(z)$ defined for all $z$.   For short distances, the scalar potential, giving rise to the Coulomb
interaction \cite{Yu2016}, is more important.   The quantum of the field is known as the scalar photon \cite{Keller-book}.
The Hamiltonian of the whole setup, $H = H_{\gamma} + H_e + H_{{\rm int}}$, is 
\begin{eqnarray}
H_\gamma &=&- s \int_{-L/2}^{L/2} dz\, \frac{1}{2}\left[ \dot{\phi}^2 + c^2 \left({\partial \phi \over \partial z}\right)^2 \right], \\
H_e &=& v_0 c_0^\dagger c_0 + v_1 c_1^\dagger c_1 + {\rm baths\ \&\  couplings},\\
H_{{\rm int}} &=& (-Q) c_0^\dagger c_0 \phi(0) + (-Q) c_1^\dagger c_1 \phi(d),
\end{eqnarray}
where $s = A \epsilon_0/c^2$ is a scale factor to give $H_\gamma$ the dimensions of energy;  $A$ is the area of the capacitor, $\epsilon_0$ is the vacuum permittivity, and $c$ is the speed of light.  
$c_0$, $c_1$, and their hermitian conjugates are fermionic annihilation and creation operators.  The 
photon field can be expressed as (in the interaction picture) 
\begin{equation}
\label{eq-phi-qantum}
\phi(z,t) = \sum_{q} \sqrt{{\hbar \over 2 \omega_q s L}}\left( a_q 
e^{i(q z-\omega_q t)} + {\rm h.c.}\right),  
\end{equation}
where $\omega_q = c |q| $  is the photon dispersion 
relation, with the wavenumber $q = 2\pi k/L$, $k$ an integer,
$a_q$ the bosonic annihilation operator of a photon of mode $q$,
and h.c.~stands for the hermitian conjugate of the preceding term. 
The scalar photon satisfies the unusual commutation relation,
$[a_q, a_p^\dagger] = - \delta_{qp}$. 
We will take the limits $L \to \infty$ and $c \to \infty$ at the end of calculation. The latter 
reproduces Poisson's equation for the field.  We should regard this limiting procedure as only a 
calculational technique; an alternative and equivalent method based on Joule heating is also possible \cite{Yu2016}.
 
Our task is to compute the energy current between the dots.  From continuity requirements of the field energy, we can establish
an expression for the current density operator to be $\epsilon_0 \dot{\phi} \partial \phi/ \partial z$.
However, to obtain a correct quantum version of the operator, we need to
symmetrize the two factors and also demand anti-normal order \cite{guidry91,jsnote1} (denoted by the bars and colons here):
\begin{equation}
j =  \frac{\epsilon_0}{2} \left[ |:\dot{\phi} {\partial \phi \over  \partial z} :| +   
|:{\partial \phi \over  \partial z}\dot{\phi}:| \right] .
\end{equation}
Anti-normal order dictates that we swap the annihilation operator to the left of the creation operator if that is not already the case.  This removes the zero-point 
motion contribution. 
We can relate the expectation value of
$j$ to the Green's functions of the photons.  The end effect of the operator ordering is to take only the positive frequency contribution of the Green's function
(a justification depends on omitting correlations between annihilation-annihilation operators, and similarly creation-creation operators, and will be presented elsewhere \cite{peng-long-PRB}).  
The average energy current per unit area at location $z$ can be obtained from 
\begin{equation}
\label{eqj}
\langle j(z) \rangle = 
\epsilon_0 \int_0^\infty { d\omega \over \pi} \hbar \omega\, {\rm Re}
{ \partial D^>(\omega, z,z') \over \partial z'}\Big|_{z'=z},
\end{equation}
where $D^{>}(\omega, z,z') = \int_{-\infty}^{+\infty} D^{>}(z,t;z',0) e^{i \omega t} dt$
is the frequency domain greater Green's function for the field $\phi$. 

We evoke the machinery of NEGF \cite{haug96,wang08review,wang14rev,aeberhard12} to calculate the required Green's functions.  First, we define the contour-ordered Green's function as
\begin{equation}
D(z,\tau;z',\tau') = -\frac{i}{\hbar} \bigl\langle T_{\tau} \Delta \phi(z,\tau) \Delta \phi(z',\tau') \bigr\rangle_{\rm noneq},
\end{equation}
where $\tau$ and $\tau'$ are Keldysh contour times, $T_{\tau}$ is the contour order operator, and the average is over a nonequilibrium steady state, $\Delta \phi(z,\tau) = \phi(z,\tau) - \langle \phi(z,\tau) \rangle_{\rm noneq}$.
The operators are in the Heisenberg picture.  Transforming into the interaction picture, and using the standard diagrammatic expansion \cite{bruus04}, we can summarize the result in a 
contour-ordered Dyson equation, which can be organized as a pair of equations in real time, the retarded Dyson equation and the Keldysh equation. 
%% Symbolically, for the 
%%Keldysh equation,  $D^{<,>} = D^r \Pi^{<,>} D^a$, here $\Pi^{<,>}$ is a sum of contributions from the nonlinear interactions at the dots.  
Due to time translational invariance, the equations become simple in the frequency domain, given as, for the Keldysh equation, 
\begin{equation}
D^>(\omega, z,z') =\!\!\! \sum_{j,k=0,1}\!\!\! D^r(\omega, z,z_j) \Pi^>_{jk}(\omega) D^a (\omega,z_k,z')
,
\end{equation}
where $z_0 = 0$, $z_1 = d$. The retarded Green's function satisfies
\begin{eqnarray}
D^{r}(\omega, z,z') &=& D_0^{r}(\omega, z,z') +\nonumber \\
&&\!\!\!\!\!\!  \sum_{j,k=0,1}\!\!\! D_0^r(\omega, z, z_j) \Pi^{r}_{jk}(\omega)
D^{r}(\omega, z_k,z'),
\label{eqDr}
\end{eqnarray}
where, $D_0^r(\omega, z,z') = - e^{i{\omega\over c}|z-z'|}/(2i \epsilon_0 A \omega/c)$,  is the free photon retarded Green's function.  The advanced Green's function is obtained
by symmetry, $D^{a}(\omega,z,z') = D^{r}(\omega,z',z)^{*}$. We also have the identity 
$D^{>}-D^{<} = D^{r} - D^{a}$. 
To make contact with the usual dyadic Green's function method \cite{Song-rev15}, one can 
turn the Dyson equation into a differential equation by operating with the
inverse of the free Green's function. However, 
due to the discrete nature of the problem, $z_j$ takes only a finite set of values. 
The above equation (\ref{eqDr}) can be solved directly, by choosing a finite
set of values $\{0,d,z,\cdots\}$.  It becomes a system of linear equations.  

In addition to the Green's functions of the photons, we also need the Green's functions of the
electrons to compute the photon self energies. A similar Dyson equation for the electrons can be established, with the
Green's function \cite{haug96}
$G_{jk}(\tau,\tau') =- \frac{i}{\hbar} \bigl\langle T_\tau  c_j(\tau) c_k^\dagger(\tau') \bigr\rangle $
and electron self-energy $\Sigma$.  The problem is completely specified if these self-energies are known.  However, for interacting systems like the electron-photon interaction $H_{{\rm int}}$, no simple closed 
form is possible (except
the formal Hedin equations \cite{hedin65}).   For the two-dot model, we present a calculation
with the self-consistent Born approximation (SCBA) \cite{bruus04}.  In this framework the photon self-energy due to the electron-photon interactions is 
essentially charge-charge correlation, in contour time, as ($j,k=0,1$)
\begin{eqnarray}
\Pi_{jk}(\tau,\tau') & = & - \frac{i}{\hbar} \bigl\langle T_{\tau} q_j(\tau) q_k(\tau') \bigr\rangle \nonumber \\
&\approx & -i\hbar Q^2 G_{jk}(\tau,\tau') G_{kj}(\tau',\tau),
\end{eqnarray}
where $q_j = (-Q) c_j^\dagger c_j$, and the second line is obtained by applying Wick's theorem.
The appearance of the self-energy $\Pi$ which is also the charge susceptibility (linear response of the induced charge by applied potential) underlines the difference between 
the present theory and the standard PvH, which relies on the current-current correlation, or frequency-dependent 
dielectric function.   The contour expression can be used to derive
the real-time formula \cite{haug96}, e.g., the retarded one in the frequency domain needed for solving the Dyson equation is 
\begin{eqnarray}
\Pi_{jk}^r(\omega) &=& -i\hbar Q^2 \int_{-\infty}^{+\infty} {dE\over 2\pi\hbar}
\Big[ G^r_{jk}(E) G^<_{kj}(E-\hbar \omega) \nonumber \\
&&  + \,
G^<_{jk}(E) G^a_{kj}(E-\hbar\omega) \Big].
\end{eqnarray}
These formulas above are general. Since the electrons cannot jump from the left lead to the right lead, 
the off-diagonal elements of the self energies are
zero; we only have
nonzero diagonal terms $\Pi_{jj}$.
The electron retarded Green's function is given by $G^r_{jj}(E) = 
1/\bigl(E - v_j - \Sigma^r_j(E) - \Sigma^r_{{\rm n},j}(E)\bigr)$, where the bath contribution to the self-energy is chosen to follow the Lorentz-Drude model \cite{wingreen-meir94}, $\Sigma^r_j(E) =  \frac{1}{2} \Gamma_j/(i + E/E_j)$, where $\Gamma_j$ and $E_j$ are the bath model constants.  The lesser Green's function is given by a Keldysh equation,
$G^<_{jj}(E) = G^r_{jj}(E) \left(\Sigma^{<}_j(E) + \Sigma^{<}_{{\rm n},j}(E)\right) G^a_{jj}(E)$.
The lesser components of the bath self energies follow from the fluctuation-dissipation
theorem of the electrons, $\Sigma^{<}_j(E) = - f_j(E) \bigl(\Sigma^r_j(E) - \Sigma^a_j(E) \bigr)$, where
$f_j(E) = 1/\bigl[\exp((E-\mu_j)/(k_B T_j)) + 1\bigr]$ is the Fermi function at temperature $T_j$ and
chemical potential $\mu_j$. 
The nonlinear self-energies (hence the subscript ${\rm n}$) $\Sigma^{r,<}_{{\rm n},j}(E)$ of the electrons arising from the Hartree and Fock diagrams under SCBA \cite{lu2007,zhanglifa2013} are (for dot $j=0$, 1):
\begin{eqnarray}
\Sigma_{{\rm n},j}^r(E) &= &i\hbar Q^2 \Big\{ -\sum_{k=0,1}D^r(0,z_j,z_k) \int G_{kk}^<(\hbar \omega) \frac{d\omega}{2\pi} + 
\nonumber \\
& & \int\frac{d\omega}{2\pi} \Big[ G_{jj}^r(E-\hbar \omega)D^>(\omega,z_j,z_j) +\nonumber \\
& &G^<_{jj}(E-\hbar \omega) D^r(\omega,z_j,z_j)  \Big]  \Big\}, \\
%% & &G_{jj}^r(E-\hbar \omega) D^r(z_j,z_j;\omega)  \Big]   \Bigg\}, \\
\Sigma^<_{{\rm n},j}(E) &=& i\hbar Q^2\!\! \int\! \frac{d\omega}{2\pi} G^<_{jj}(E \!-\! \hbar \omega) D^<(\omega,z_j,z_j).
\end{eqnarray} 
The current formula, Eq.~(\ref{eqj}), can be further simplified using the solution of the Dyson equation and then taking
the limit $c\to \infty$, given
\begin{equation}
\langle j \rangle A = \int_0^{\infty}\!\! { d\omega \over \pi} \hbar \omega
|D_{01}|^2 i \bigl( 
  \Pi_{11}^{>} {\rm Im} \Pi_{00}^r
- \Pi_{00}^{>} {\rm Im} \Pi_{11}^r
\bigr),
\end{equation}
where $D_{01} = \bigl[\Pi_{00}^r \Pi_{11}^r/C - (\Pi_{00}^r + \Pi_{11}^r)\bigr]^{-1}$; $C=\epsilon_0 A/d$ is the capacitance.

\begin{figure}
\includegraphics[width=\columnwidth]{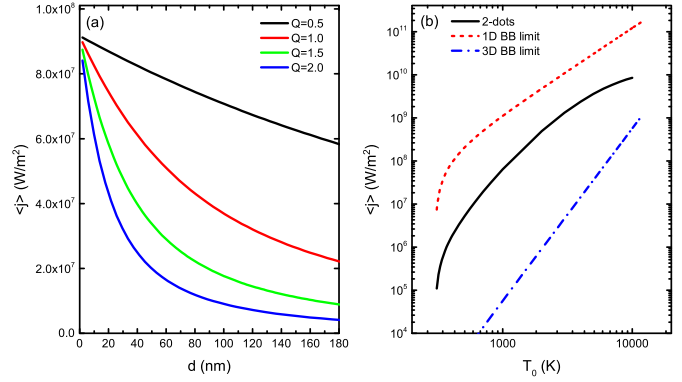}%
\caption{(a) Heat current density $\langle j\rangle$ as a function of distance $d$ with different maximum charge $Q$ for the two-dot model. We set the temperatures of the baths at $T_0=1000\,$K, $T_1=300\,$K, chemical potentials $\mu_0=0\,$eV, $\mu_1=0.02\,$eV, and 
onsite $v_0=0$, $v_1=0.01\,$eV, area $A=389.4$ (nm)$^2$,  and the electron bath parameters $\Gamma_0 = 1\,$eV, $\Gamma_1=0.5\,$eV,
$E_0 =2.0\,$eV, $E_1 = 1.0\,$eV.
(b) The temperature dependence of heat current density. Here, we set $T_1=300\,$K and vary $T_0$, with $d=35.5\,$nm and $Q= 1\, e$. Other parameters are the same as for Fig.~2(a). }
\end{figure}

%% discuss results
We now discuss the results of the two-dot model.
Figure 2(a) shows the heat current density $\langle j \rangle$ as a function of the two-dot separation $d$. 
A key parameter is the area $A$; we choose it to be close to the experimental values
in Ref.~\cite{Kloppstech2015}.  The results are insensitive to the onsite potentials $v_j$ and
chemical potentials $\mu_j$, provided $\Gamma_j$ is large.
We note that the energy current takes an exact scaling form, $\langle j \rangle A = F(x)$, $x =A/(Q^2d)$,
with the area $A$, distance $d$, and charge $Q$.  Further analysis shows that $F(x) \propto x^2$ for small
$x$ and approaches a constant for large $x$.   Under the strong coupling limit, $k_B T_j \ll \Gamma_j$, in addition to
a $T_j^4$ temperature dependence, the current is proportional to $A/(d^2 Q^4)$ for small $x$ and 
$1/(A \Gamma^2)$ for large $x$ [here $\Gamma \approx \max( \Gamma_0, \Gamma_1)$].  The crossover from one type of behavior to the other type of behavior
is controlled by 
$\Gamma \approx Q^2/(2C)$.
Alternatively, a length scale can be obtained from $D_{01}$, giving
$\tilde d =  - \epsilon_0 A [1/\Pi^r_{00}(0)+ 1/\Pi^r_{11}(0)]$.   The current decreases as 
$1/d^2$ for large $d$ and saturates at the scale given by $\tilde d$ (about $40\,$nm with our choice of 
parameters) for small $d$.

This phenomenon is different from near-field radiative heat transfer results at a length scale
less than $\hbar c/(k_B T)$ dominated by evanescent modes when the 
transverse component of the wave vector 
$q_\perp > \omega/c$.  There are no 
evanescent modes in our 1D model since
$q_\perp = 0$ permanently. 
Besides, Figure 2(a) also shows large values of heat transfer.
% due to  surface charge resonance which is not obtained in the standard fluctuational electrodynamics.  
At the saturation value, the heat current density is approximately $9 \times 10^7$ W/m$^2$, which is about a thousand times larger than the black-body (BB) limit of $5.6\times 10^4$ W/m$^2$.    This enhancement 
%over BB value of a thousand 
is at least one order of magnitude larger than a typical PvH theory prediction for metals, which is in the range of the hundreds. 
Compared to a one-dimensional Landauer formula (1D BB) result with perfect transmission, i.e. $1.1\times 10^9$ W/m$^2$, our numbers are about 1/12-th of that upper limit \cite{Abdallah10}. 
Such enhancement is mainly due to transverse confinement (there is
only one transmission mode) and the small area $A$.
The temperature dependence of the current density is plotted in Fig.~2(b). 
Asymptotically for large $T_0$ fixing $T_1$,
the Stefan-Boltzmann law gives the fourth power of $T_0$ and the 1D BB limit gives a quadratic function of $T_0$.  
%The quantum dot model demonstrates
%an unusual temperature dependence which could be related to the specific density of states of the quantum 
%dots as compared to bulk systems.

The results presented above are for strong couplings, where $\Gamma_j$ is comparable to electron energy scale of
order eV.  For weak couplings, SCBA convergence is difficult.   However, we observe the physics remains
qualitatively the same if we do not do  self-consistency.  In such a framework, the photon self-energies
$\Pi_{jj}^>$ can be approximated by the fluctuation-dissipation theorem, assuming each dot is in local thermal
equilibrium.  Within this framework of approximation, we can recover a Caroli/Landauer formula \cite{peng-long-PRB}.

%%% summary part and perspective
In summary, we have presented a fundamental theory of near-field
heat transfer due to Coulomb interactions of the electrons that applies to distances approaching atomic lattice constants.  
This is different from the usual PvH theory where only  the transverse, radiative field is considered. 
At ultra-short distances, we should consider charge fluctuations and Coulomb interactions through a 
scalar field, which give an additional channel of heat transfer not contained in the Poynting vector [Note that if
the vector potential ${\bf A}=0$, magnetic induction ${\bf B}$ is zero, so the Poynting vector is zero]. 
The contributions from the scalar potential $\phi$ are not small at short distances  \cite{Yu2016} as compared to the usual PvH 
results which include only the Poynting vector term involving the vector field. 
The approach proposed here opens the way for the treatment of other geometries such a surface and a tip. 
Such calculations could resolve the controversies regarding the recent experiment \cite{Kloppstech2015}. 
Our approach can be interfaced with first principle calculations, thus enabling more rigorous predictions of near-field properties.

The authors thank Lifa Zhang for stimulating discussions and Han Hoe Yap for pointing out an error in an earlier version of the paper. This work is supported by FRC grant R-144-000-343-112 and MOE grant R-144-000-349-112. 

% Create the reference section using BibTeX:
\bibliographystyle{eplbib}
\bibliography{capacitor-NF}

\begin{thebibliography}{10}
\bibitem{jsnote1}
The anti-normal order is because the creation operators annihilate the photon
  vacuum state, $a^\dagger_q | 0 \rangle=0$.
\end{thebibliography}

\end{document}